# 3D Cosmic Ray Muon Tomography from an Underground Tunnel

Elena Guardincerri[1]*, Charlotte Rowe[1], Emily Schultz-Fellenz[1], Mousumi Roy[2], Nicolas George[2], Christopher Morris[1], Jeffrey Bacon[1], Matthew Durham[1], Deborah Morley[1], Kenie Plaud-Ramos[1], Daniel Poulson[1,2], Alain Bonneville[3], Richard Kouzes[3]

**Affiliations:**

[1]Los Alamos National Laboratory

[2]University of New Mexico

[3]Pacific Northwest National Laboratory

*Correspondence to: elena.guardincerri@lanl.gov

**Abstract**: We present an underground cosmic ray muon tomographic experiment imaging 3D density of overburden, part of a joint study with differential gravity. Muon data were acquired at four locations within a tunnel beneath Los Alamos, New Mexico, and used in a 3D tomographic inversion to recover the spatial variation in the overlying rock-air interface, and compared with *a priori* knowledge of the topography. Densities obtained exhibit good agreement with preliminary results of the gravity modeling, which will be presented elsewhere, and are compatible with values reported in the literature. The modeled rock-air interface matches that obtained from LIDAR within 4 m, our resolution, over much of the model volume. This experiment demonstrates the power of cosmic ray muons to image shallow geological targets using underground detectors, whose development as borehole devices will be an important new direction of passive geophysical imaging.

**One Sentence Summary:** 3D density modeling using an underground cosmic ray muon detector points the way to new borehole methods for passive shallow geophysical imaging.

**Introduction**

Cosmic-ray muons are naturally produced in the atmosphere by the interactions of primary cosmic rays with the nuclei present in the atmosphere itself. They reach the surface of the Earth at a rate of about 1/cm2/min with a well-known, broad, continuous energy distribution and a mean energy of ~4 GeV (*1*). The attenuation of cosmic ray muons passing through matter can be used to estimate the density of objects through which they pass. Compared to background muon flux rates, the fraction of muons detected along a given path provides the integrated density length along that path.

The first three-dimensional density image of a geophysical object using cosmic-ray muons (*2, 3*) imaged a volcano using low-angle muon data from a detector located at multiple positions around it. Later, the same team obtained a tomographic image of a volcano combining muon attenuation data from two different angles with gravity measurements (*4*), using a topographic map of the volcano as a prior. Some applications with deployment of detectors in tunnels have also been conducted in ore exploration (*5*) and geologic mapping (*6*). We present muon tomography of the Los Alamos mesa obtained using a detector placed at multiple positions within a tunnel beneath the target overburden. Our 3D inversion recovers well the overlying air/rock interface and the bulk density of the overburden, without invoking any prior constraint on the topography of the mesa.

**Experiment Site**

Our data were acquired at four locations within a tunnel, excavated horizontally under the main Los Alamos town-site mesa (*7*) (Figure 1). The tunnel has a length of about 100 m and an overburden ranging between 0 m (at the entrance) and 100 m at its innermost point. The long axis of the tunnel is oriented N5°W. Now decommissioned and de-classified from its Cold War status, the tunnel was used in this study to generate a tomographic density image of the mesa by inverting the measured attenuation of cosmic-ray muons through the overburden.

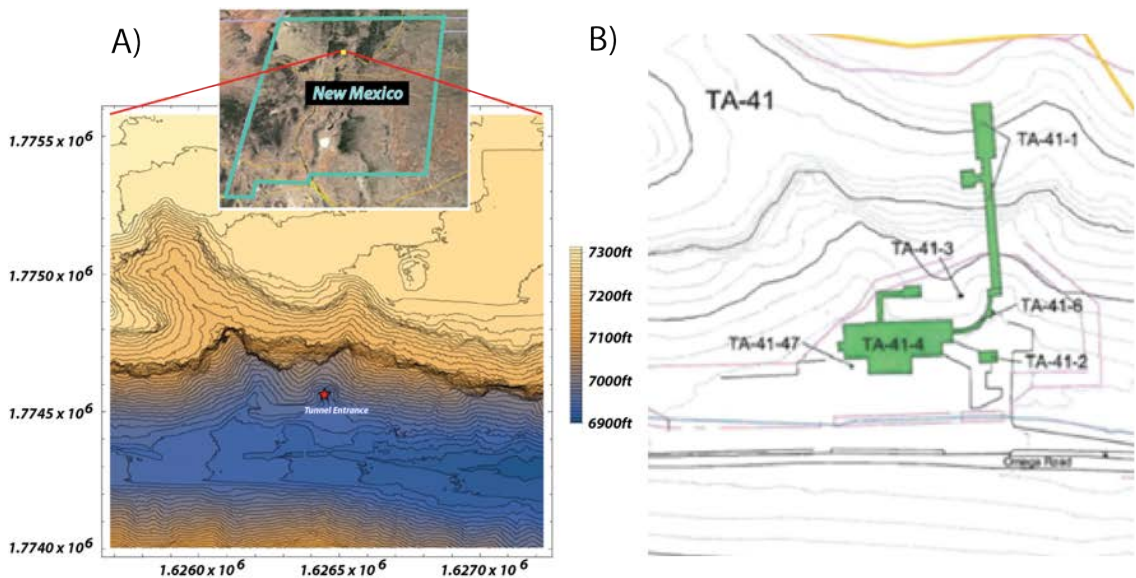

Figure 1: Field experiment site and layout. A) Map showing position within New Mexico (inset) of the field area and a topographic map of the Los Alamos Canyon site with tunnel entrance marked by a red star. Contour intervals are in feet above sea level. B) plan view of the tunnel.

The mesa is comprised of the Quaternary Bandelier Tuff, a sequence of ash-flow tuffs deposited during the most recent major caldera-forming eruptions of the nearby Jemez volcano, located approximately 10 km west of the site. Figure 1a is a location map (inset) for the target with detailed topography of Los Alamos Canyon and the mesa above it. The tunnel entrance is indicated by a red star. Figure 1b illustrates the configuration of the tunnel and ancillary structures.

At the tunnel site (Figure 2) the Tshirege Member (upper unit) of the Bandelier Tuff is exposed (*8,9*). This member can be further subdivided into cooling units, which have demonstrable mineralogic and physical variations due to the episodic nature of the Tshirege Member eruptive sequence.

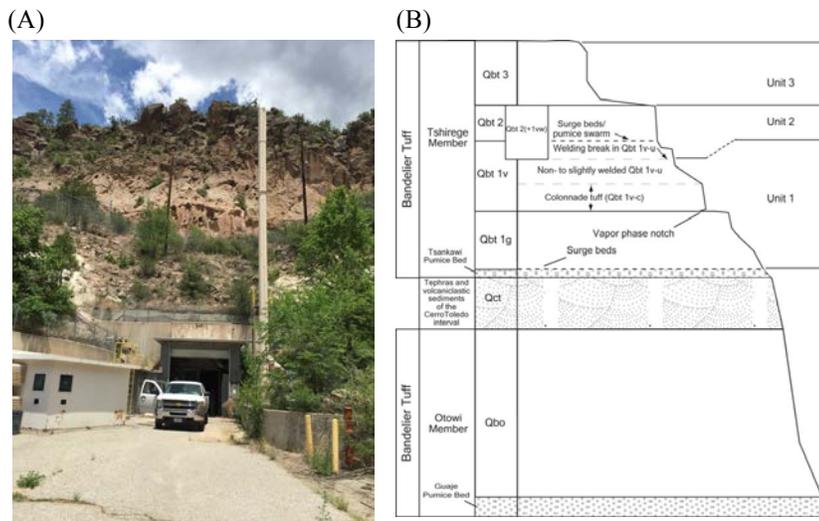

**Figure 2:** Field site exposure and stratigraphic column. A) Tunnel entrance and first tier of overburden B) Stratigraphy of Bandelier Tuff (*8*). Visible here is only Unit 1 of the Tshirege Member, overlying the uppermost Ottowi. Stratigraphy in b) has been aligned with exposed units visible in a).

**Method**

The processes contributing to the muon attenuation in matter are well known (*10*), so that a measurement of muon attenuation can be used to determine the amount of matter, or "range" R (density times length) traversed by the particles along their path. If $R_j$ is the range of a muon travelling along a path j, then

$$R_j = \sum_i L_{ji} \rho_i \quad (1)$$

where each $L_{ji}$ is the path length of muons along path j through the i-th matter element, and $\rho_i$ is the density of the i-th matter element. Equation (1) is leveraged to determine the three-dimensional density distribution of an object from the range of muons traveling along different paths through it, via an inverse linear problem.

Both the stopping power and the range R (density times length) for muons have been calculated and tabulated as a function of the initial energy of the muons for different materials and compounds (*11*). The relationship between the minimum energy $E_{min}$ that a muon must possess to traverse a material without being absorbed and its range R is tabulated for a set of elements and materials:

$$E_{min} = E_{min}(R) \quad (2)$$

If $f_\mu = f_\mu(\theta, E)$ is the energy distribution of cosmic-ray muons at the Earth's surface as a function of their zenith angle θ and of their kinetic energy E, then the ratio of the number of muons $N_{surv}$ surviving the passage through an object to the number $N_{tot}$ of incident muons is given by:

$$\frac{N_{surv}}{N_{tot}} = \frac{\int_{E_{min}}^{\infty} f(\theta, E)}{\int_{0}^{\infty} f(\theta, E)} \quad (3).$$

Equations (2) and (3), together with the knowledge of f(θ,E) can be used to obtain the range R of the muons along a path from their measured attenuation. We used values tabulated (*11*) for standard rock to model the dependence $E_{min}(R)$ and the probability distribution functions provided by the Cosmic-Ray Shower Monte Carlo software (*12*) to model the energy and angular distribution of cosmic-ray muons at the 2100 m elevation of Los Alamos.

**Data acquisition and analysis**

The Los Alamos National Laboratory (LANL) Mini Muon Tracker (MMT) (*13*), shown in Figure 3, was used for this experiment. The detector consists of two modules made of 576 sealed, aluminum drift tubes arranged in planes. Each of the two modules consists of six horizontal planes of tubes; tubes in each plane are oriented orthogonally to those of its immediate neighbor. The detector is capable of tracking muons with 2.5 milliradians angular resolution across a surface of 1.4 m$^2$.

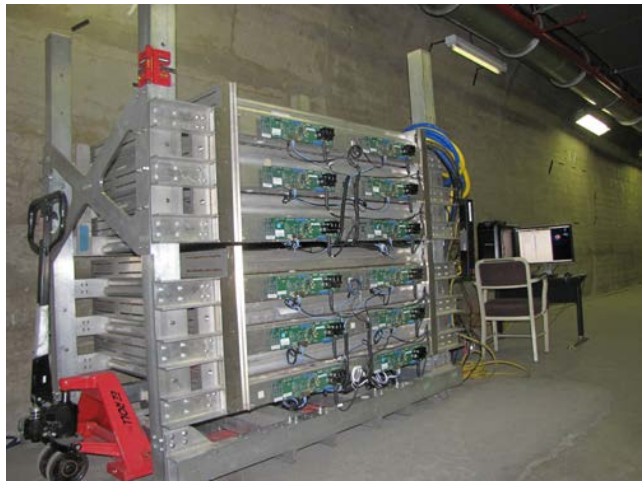

Figure 3: The Mini Muon Tracker (MMT) deployed inside the tunnel.

Table 1 presents the data acquired at each of the detector positions in the tunnel, as well as outside for obtaining true background flux.

| Position # | Distance (m) from Portal | Hours of Data | Muons per hour |
|---|---|---|---|
| 0 | 40.2 | 392 | 11,750 |
| 1 | 77.82 | 564 | 7,420 |
| 2 | 31.57 | 331 | 41,800 |
| 3 | 4.95 | 433 | 159,358 |
| Outside | N/A | 348 | 403,766 |

**Table 1:** Muon Data Acquisition

Muon tracks obtained from the data at each location were divided in 968 angular bins each one having $\Delta\theta = \Delta\varphi = \pi/44$ in the range $\theta \in [0, \pi/4]$, $\varphi \in [-\pi,\pi]$. As a result, 968 x 4 = 3872 muon paths were considered.

The detector's acceptance is limited to a solid angle of ~ 45 degrees about the vertical axis and becomes marginal for larger inclination angles. Since we used the ratios (see Equation 3) between the number of muons recorded underground and the number of muons recorded on the surface in each angular bin, any detector acceptance artifacts cancel and do not affect the data. Figure 4 shows three views of the mesa with the four detector locations and the 45 acceptance cones above them.

To solve equation (1) we parameterized the $5.4 \cdot 10^6$ m$^3$ volume above the tunnel as a rectilinear model comprised of 84,672 cubic cells with 4 m sides. Cells residing outside of the detector's acceptance at all locations were excluded from the inversion to reduce the computational time required to solve the problem; hence, the number of cells considered was 38,546. Equation (1) cannot simply be inverted: the matrix L is singular since this problem is underdetermined in part of the parameter space considered. We therefore apply linear regularization.

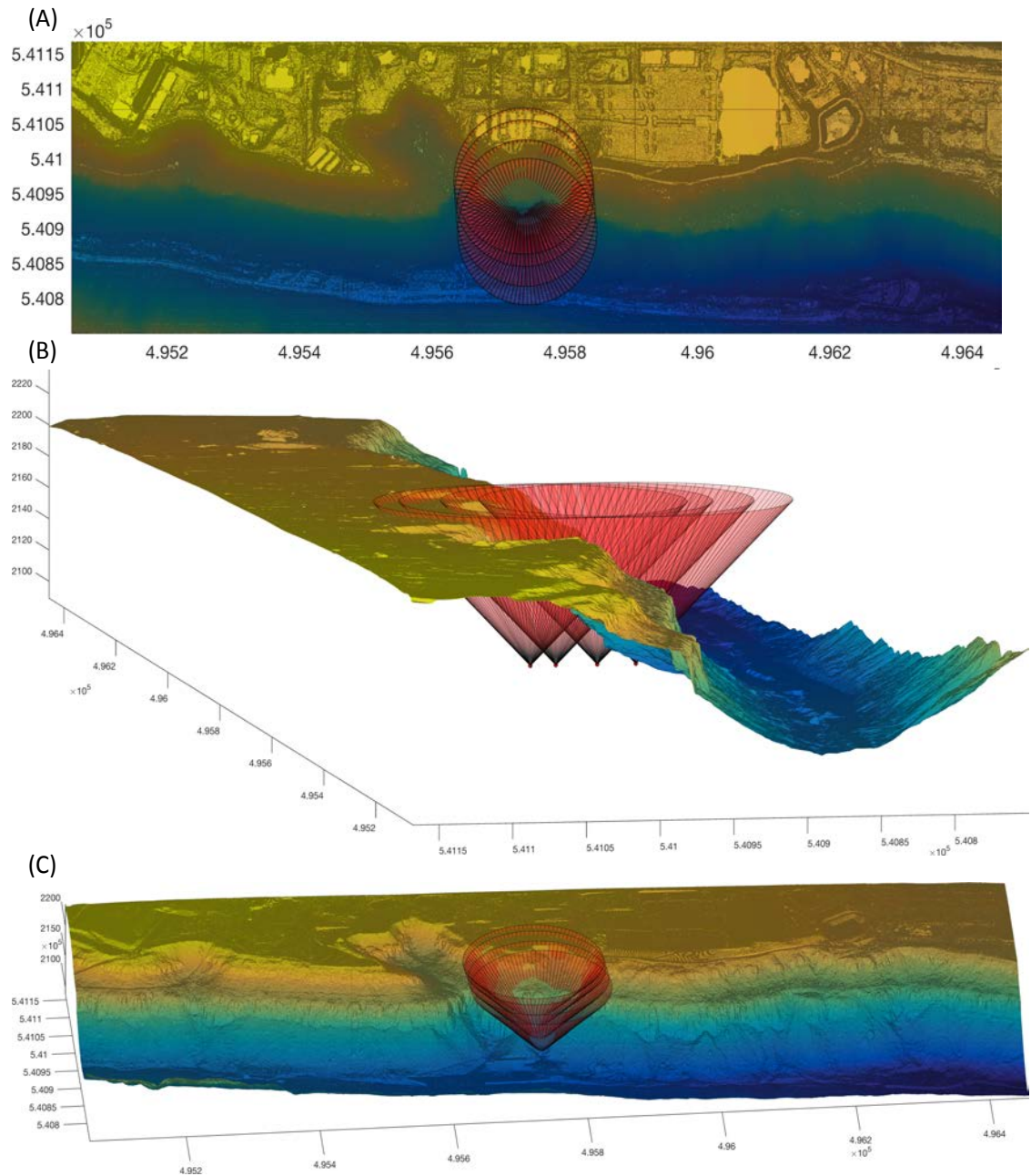

Figure 4: Muon acceptance cones for 4 positions of the detector. Red cones indicate limits of the MMT acceptance for muon tracks. A) Map view. B) View along strike of canyon. C) View perpendicular to strike of canyon. Axes labeled in NAVD88 coordinates. Figure courtesy of Megan Lewis.

Different regularization methods exist (*14*); here we adopt a smoothing constraint on the rock density through an exponential covariance (*16*):

$$C_{reg}(i,j) = \sigma_\rho^2 e^{-|r_i-r_j|/\lambda} \quad (4)$$

where $\sigma_\rho$ is the *a priori* error on the density, $\lambda$ is a correlation length and $|r_i-r_j|$ is the distance between the i-th and the j-th cells. A generalized $\chi^2$ can then be defined as:

$$X_{gen}^2 = X_d^2 + X_{reg}^2 = (R-R_o)C_d^{-1}(R-R_o) + (\rho-\rho_0)C_{reg}^{-1}(\rho-\rho_0) \quad (5)$$

where R is given by Equation (1), $R_0$ are the density lengths obtained from the data using Equation (2) and Equation (3), $\rho_0$ is an initial guess on the density of each cell and $C_d$ is a diagonal matrix whose (j,j) entry is the standard deviation of the muon range along path j based on the statistics collected along that path. Equation (5) can be minimized (*15*) with the constraint from equation (1) and the solution is:

$$\rho = \rho_0 + C_{reg} \cdot L^T \cdot (C_d + L \cdot C_{reg} \cdot L^T)^{-1} \cdot L \cdot C_{reg} \quad (6)$$

The solution of Equation (6) depends on three parameters, $\rho_0$, $\sigma_\rho$ and $\lambda$. In order to optimize them we used synthetic data.

We obtained a LIDAR map of the Los Alamos mesa to reproduce its contours and assigned a uniform density of 1.8 g/cm³ to the rock, and 0 g/cm³ to the air. We then employed a forward model to calculate the expected muon rates and angular distributions at the actual detectors locations, and subsequently applied the inversion algorithm described above.

The value of $\rho_0$ was chosen equal to 1.5 g/cm$^3$, the average value for the cells in our density model. The values of $\sigma_\rho$ and $\lambda$ were chosen in order to maximize the agreement between the density model used and the result while keeping the ratio $X_d^2$/NDF close to 1, NDF being 3872, the number of degrees of freedom for the un-regularized problem:

$$\begin{cases} \sum_{i=1}^{N_{vox}} |\rho - \rho_{synth}|^2 = \min \\ X_d^2 / NDF = ((R - R_0) C_d^{-1} (R - R_0)) / NDF = 1 \end{cases} \quad (7)$$

The optimal values of $\lambda$ and $\sigma_\rho$ were found respectively equal to 188 m and 1.2 g/cm$^3$. For these values $X_d^2$/NDF = 1.001 and $\sum_{i=1}^{N_{vox}} |\rho - \rho_{synth}|^2$ = 22,862 (g/cm$^3$)$^2$. Figure 5 shows the dependence of the solutions on the two parameters $\lambda$ and $\sigma_\rho$ when $\rho_0$ = 1.5 g/cm$^3$, and in particular shows that the solution is quite stable with respect to variations of $\lambda$ and $\sigma_\rho$.

Figure 6 shows how the quantity $\sum_{i=1}^{N_{vox}} |\rho - \rho_{synth}|^2$ depends on the distance of the cells considered to the straight line running along the center of tunnel, where the muon detector was deployed. Muon trajectories often did not cross in those cells along the tunnel or immediately around it, so that the agreement between the input model and the solution obtained from it is sub-optimal for regions along, and close to, the tunnel.

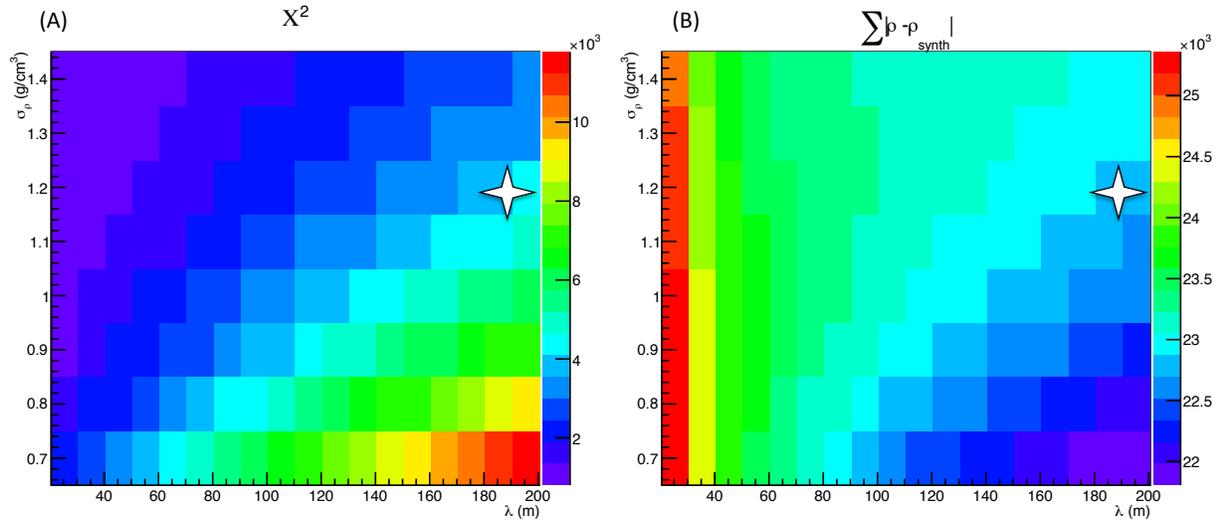

Figure 5: Solution dependence on regularization and density parameters. A) Dependence of $X_d^2$ on $\lambda$ and $\sigma_\rho$ for $r_0 = 1.5$ g/cm³. B) Dependence of $\sum |\rho - \rho_{synth}|^2$ on $\lambda$ and $\sigma_\rho$ for $r_0 = 1.5$ g/cm³. The white star indicates the values of $\lambda$ and $\sigma_\rho$ that satisfy Equations (7).

When more distant cells are considered, the agreement improves and reaches an optimal value for distances in the neighborhood of 65 m from the tunnel. For larger distances, the trajectories of muons become increasingly parallel to one another, thus unable to resolve density anomaly positions.

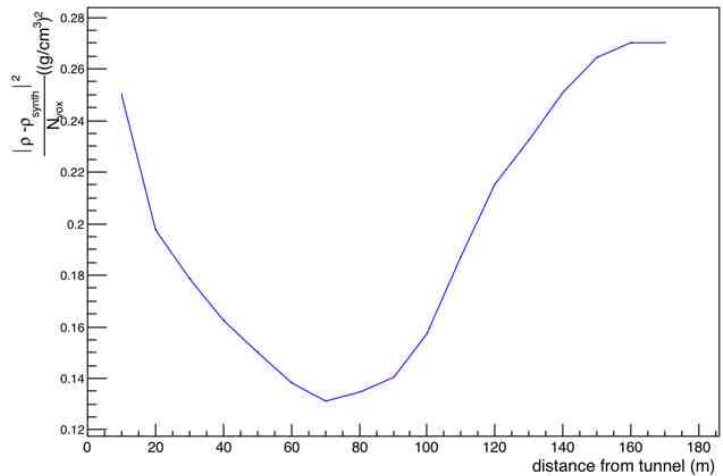

Figure 6: Dependence of solution quality on distance of cells from tunnel axis. $\frac{\sum_{i=1}^{N_{vox}} |\rho - \rho_{synth}|^2}{N_{vox}}$ for cells at distances from a line running along the center of the tunnel. Only those cells inside a cylinder of given radius r, shown on the X axis, were used to calculate the quantity shown on the Y axis.

**Results**

The data were fitted using the method described above, with $\rho_0$ = 1.5 g/cm³, $\sigma_\rho$=1.2 g/cm³ and $\lambda$ = 188 m. We present the results in Figure 7. Only those cells having density larger than 0.4 g/cm³ and inside the detectors' acceptance are shown for clarity. Cross-sections and map views are compared against the corresponding LIDAR values. The profile of the mesa is clearly visible in the tomographic results, and the average value obtained for the rock densities of (1.2±0.9) g/cm³ is compatible with the range [1.28 g/cm³, 1.84 g/cm³] for the density of rhyolitic tuffs based on the measurements by *(16)*. The average density determined for air cells is (0.0±0.3)g/cm³. For the purpose of calculating this value, a cell's assignment to rock vs. air was made by comparing the density profile determined inverting the muon data against the threshold value of 0.4 g/cm³. Topographic contour maps of the mesa for our model vs. LIDAR appear in Figures 7E and 7F. Figure 8 shows the difference between the elevation contours obtained respectively from the inversion of the muon data and from the LIDAR scan. The difference is, over most of the range within the detector's acceptance, smaller than 4 m (the size of the cells, our intrinsic resolution). It becomes larger where the cliff is steeper, and therefore small offsets in the horizontal direction can produce large errors in the vertical direction, and at the edge of the acceptance region, where the density of the muon tracks is lower and the problem is less constrained.

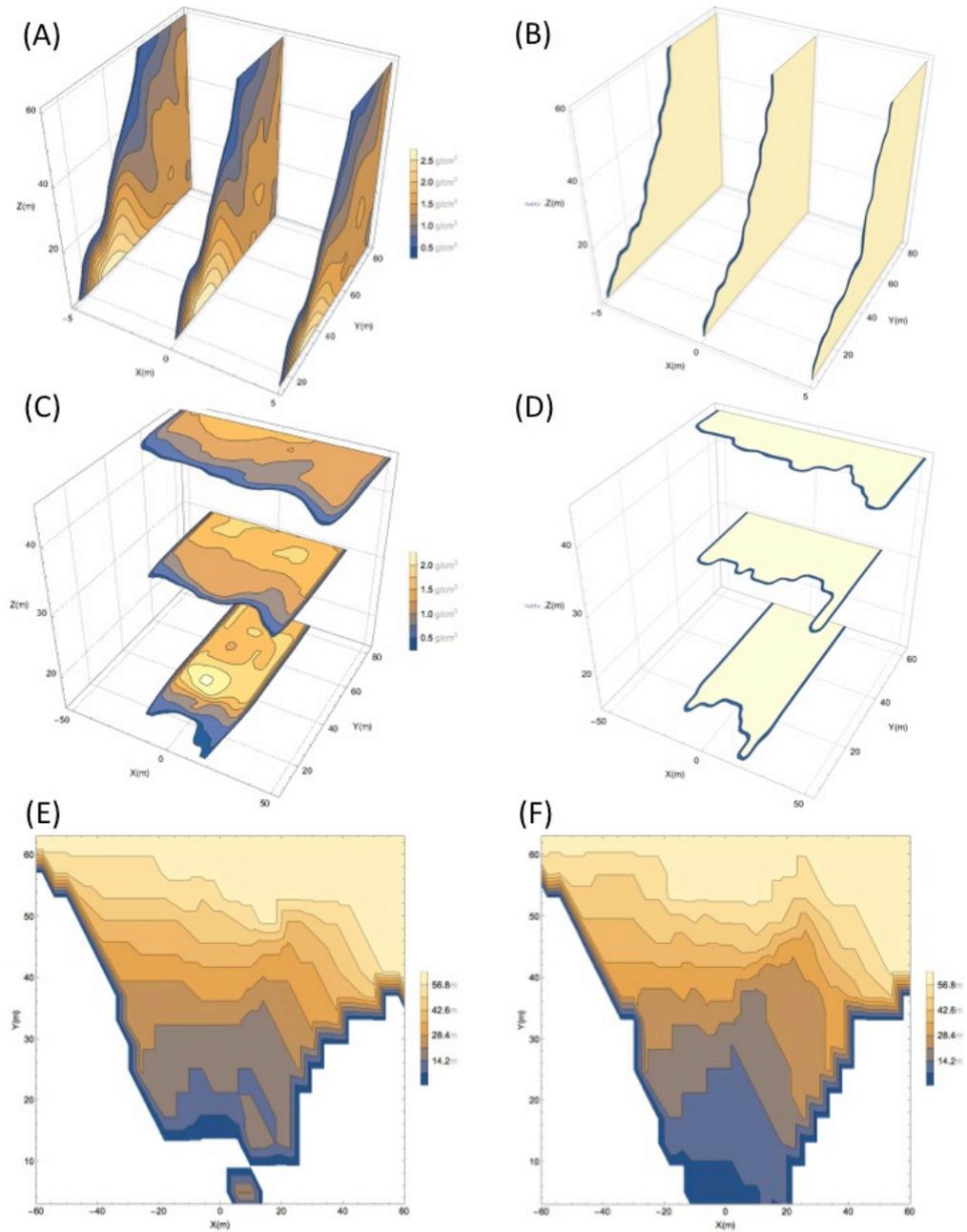

Figure 7: Tomographic results and comparison to LIDAR rock/air interface. Areas outside detector's acceptance are not shown. (A) and (B) Vertical cross section through model and LIDAR, respectively. (C) and (D) Horizontal cross sections through model and LIDAR, respectively. (E) and (F) Model estimation, and LIDAR validation, respectively, for topographic elevation contours in map view.

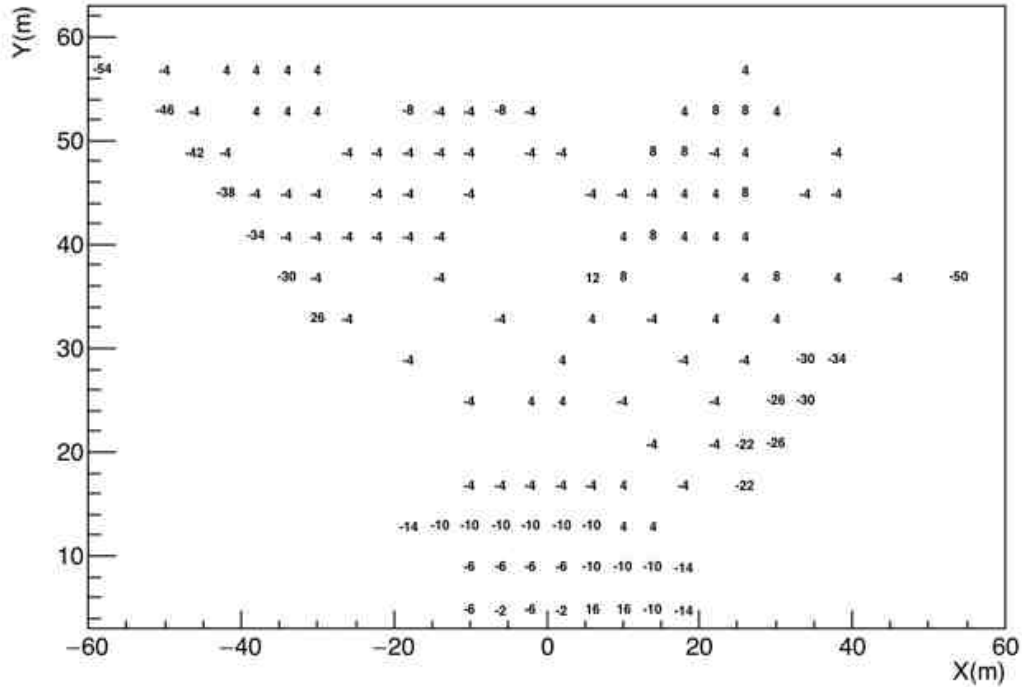

Figure 8: Difference between the elevation contours obtained from the actual muon data, shown in Figure 7(E) and 7(F)

**Conclusions**

We demonstrate 3D tomography of a geological structure obtained by an unconstrained inversion of cosmic ray muon data acquired underground. The image reproduces well the profile of the overburden, and the density obtained for its interior is compatible with independent estimates from gravity as well as standard published densities for this lithology.

Data were acquired at four locations along a straight line within the tunnel, mimicking the configuration of a string of small borehole detectors in a horizontal borehole beneath a target reservoir or other body; we thus demonstrate the potential of borehole muon geophysics for imaging subsurface targets. A prototype borehole detector was developed in this project and tested in the tunnel; its measurements, although of somewhat lower resolution, show fidelity to those from the MMT (17) and prove promising for this nascent technology.

**Acknowledgments**

This work was supported by the Department of Energy (DOE) Subsurface Technology and Engineering Research, Development, and Demonstration program and by the LANL Center for Space and Earth Science. We used resources provided by the Open Science Grid (*18, 19*), which is supported by the National Science Foundation and the U.S. DOE Office of Science. We thank Megan O. Lewis for kindly providing the images in Figure 4. This is Los Alamos Publication LA-UR-16-XXXXXX.